\documentclass[aps,prb,showpacs,twocolumn]{revtex4-1}
\usepackage{amssymb}
\usepackage{amsmath}
\usepackage{graphicx}
\usepackage{epsfig}
\usepackage{subfigure}

\begin{document}

\title{Robustness of the pumping charge to dynamic disorder}
\author{R. Wang, and Z. Song}
\email{songtc@nankai.edu.cn}
\affiliation{School of Physics, Nankai University, Tianjin 300071,
China}

\begin{abstract}
We investigate the effect of disorder on a gapped crystalline system by
introducing a class of local quantities for an energy band, which is
referred as to band correlation function (BCF) and is the sum of correlation
functions for all eigenstates of the band. We show that the BCFs are robust
in the presence of disorder if the band gap is not collapse. The eigenstate
set of an energy band can be almost completely mapped onto the perturbated
eigenstate set, referred as to quasi-closed mapping, when it is sufficiently
isolated from other bands. Some features relate to translational symmetry
may emerge in a randomly perturbed system. We demonstrate this by simulating
numerically the pumping process for a 1D Rice-Mele (RM) model with disorders
on the hopping strength and on-site potential. It is shown that the
quantized pumping charge is robust against with the dynamic disorder. This
result indicates the possibility of measuring the topological invariant in
experimental system with imperfection.
\end{abstract}

\maketitle

\section{Introduction}

The notions of topology have been invoked in condensed matter physics and
material sciences since the connection of integer quantum Hall conductances
with topological Chern invariants was discovered\cite{DJ0}. Topological
theory has been well established in condensed matter physics \cite%
{Kitaev,RyuPRL2002,Greiner,SCZhang,Bernevig06,Kane,LFuPRB,LFuPRL,Ludwig,RyuNJP,KaneRMP,XLQi,Xu,Burkov,Young,Wang,Wang1,Bardyn2012,Esslinger,Weng,LLuScience,CTChanNP,Liu,Leykam2,Ryu,Kunst2018,Armitage}%
. So far, the topological feature in matter is mainly described by
topological invariant, Chern number, which greatly expands our knowledge on
state of matters. It has explicit definition for crystalline system, in
which the momentum is a crucial quantity, and thus the translational
symmetry is necessary. On the other hand, it is usually claimed that the
topological property is robust against disorder perturbations. This can be
manifested in a rigorous manner by the existence of edge states of the
system immune in the presence of disorder. The robustness of topological
edge states has been actively explored in a wide variety of quantum Hall
systems, topological insulators, and topological superconductors \cite%
{KaneRMP,XLQi}. A natural question thus arises as to whether the topological
feature is insensitive to disorder in an approximate manner. It is a
different and less explored subject but has potential applications in
realistic systems. For instance, a Hall conductivity, that is related to the
first Chern number for perfect topological system, appears as approximate
integer in the presence of imperfection. Recently, some pertinent
theoretical effort has been devoted to the topological phases in
noncrystalline lattices \cite{I.C,Zohar,Adhip}.

In this work, we study the influence of disorder perturbation on the
topological properties of a crystalline system. We show that there exist
global quantities which remain unchanged under the disorder perturbation if
the energy gap is not collapsed. It provides a unified framework to discuss
topological invariant for disorder-perturbed system, such as Rice-Mele\ (RM)
model with random distribution of hopping strength and on-site potential. We
focus on a similar concept with Hall conductance in $2$D topological system,
the Thouless\ pumping charge \cite{D.J,Y.H2}. It is a local quantity, which
can be measured without the need of translational symmetry. To demonstrate
this point, we study a quasi-adiabatic process by numerical simulation. We
implement the dynamic disorder by discretized time evolution under a
sequence of Hamiltonians based on a set of time-dependent random numbers.
Both the analytical analysis and numerical simulation show that pumping
charge, as a measurable Chern number, is robust against the disorder. Our
work provides a way of extracting topological invariant from an imperfect
system.

This paper is organized as follows. In section \ref{Robustness of band
correlation function}, we present the concept of quasi-closed mapping and
the related application. In section \ref{Disordered RM model}, we
investigate the connection between the crystalline and noncrystalline RM
models analytically and numerically. In section \ref{Robustness Of Pumping
Charge}, we present the numerical results about the pumping charge in the
presence of disorder.\ Section \ref{Summary} summarizes the results and
explores its implications.

\section{Robustness of band correlation function}

\label{Robustness of band correlation function}
\begin{figure}[tbph]
\includegraphics[ bb=67 133 506 679, width=0.49\textwidth, clip]{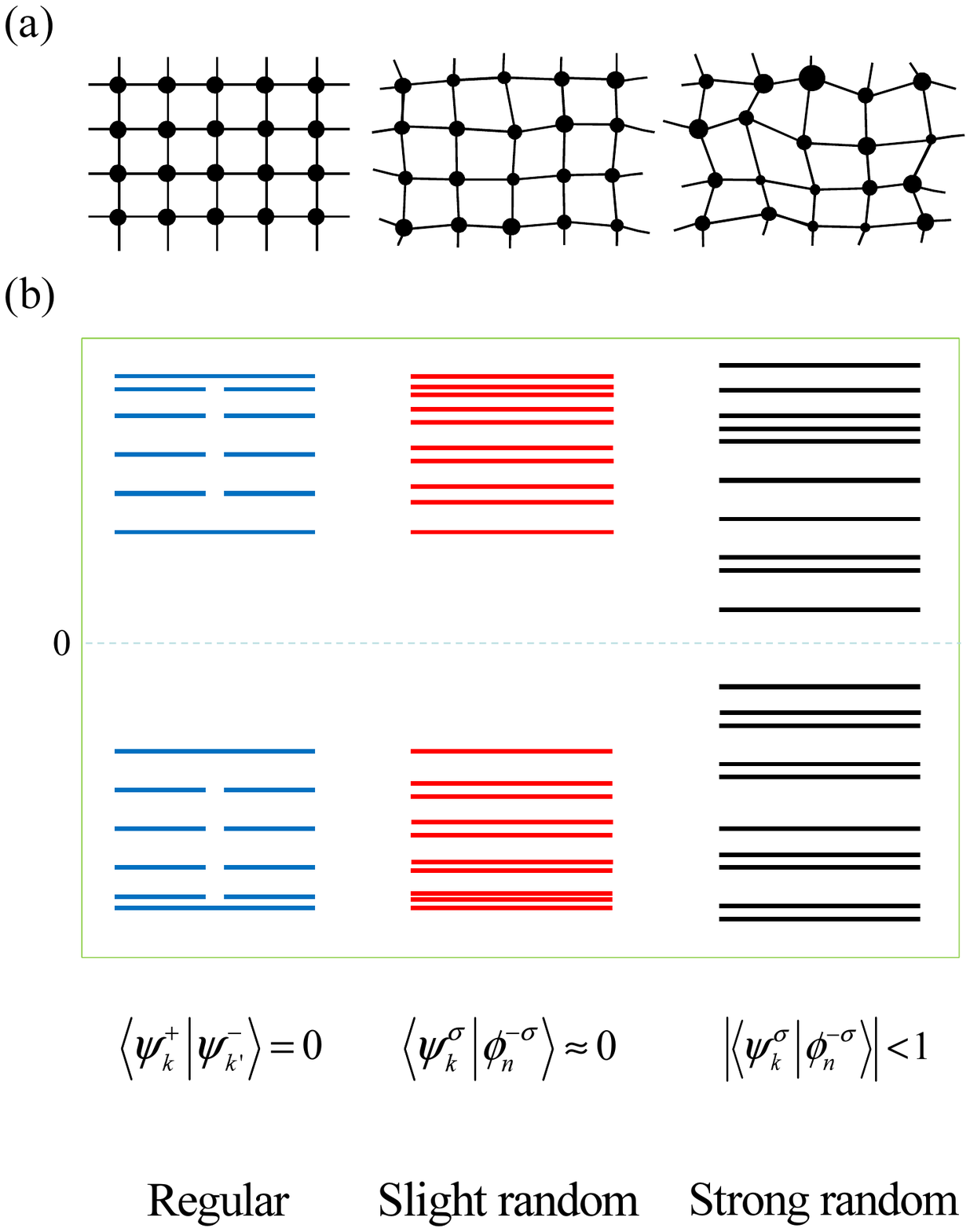}
\caption{(Color online) Schematic illustration of the concept of
quasi-closed mappings. (a) Schematics of lattice systems with different
imperfections: regular lattice with translational symmetry, noncrystalline
lattices arising from slight and strong random parameters. (b) The
corresponding spectra and the implication of perfect invariant sub-space
(left) and quasi-closed mapping (middle). Slight disorder may destroy the
degeneracy in the left spectrum but maintains structure of spectrum
approximately, while in the right spectrum, strong disorder hybridizes the
energy levels between upper and lower bands of the original system.}
\label{fig1}
\end{figure}

The concept of energy band
originates from the periodic potentials is always associated with
translational symmetry. However, a slight disorder perturbation may
hybridize the energy levels but cannot destroy the density of states and
collapse the energy gap. In this section, we investigate what remains of the
energy band in the presence of disorder. We consider a generic Hamiltonian
\begin{equation}
H=\sum_{i,j=1}^{N}\gamma _{ij}c_{i}^{\dag }c_{j}+\mathrm{H.c.},
\end{equation}%
which describes fermions hop from one site to another with amplitude $\gamma
_{ij}$, and the on-site energy $\gamma _{ii}$. Here $c_{i}^{\dag }$ denotes
fermion operator. We start with the case with
\begin{equation}
H=H_{0}+H^{\prime },
\end{equation}%
where $H_{0}$ has translational symmetry. For simplicity, we assume that
there are two energy bands, with the single-particle eigenstate set $\left\{
\left\vert \psi _{k}^{\sigma }\right\rangle \right\} $ with $\sigma =\pm $
and $k\in \lbrack 1,N/2]$. Here $k$\ labels the energy levels with wave
vector $(k_{x},k_{y},k_{z})$ if a 3D system is considered. Obviously we have
$\langle \psi _{k}^{+}\left\vert \psi _{k^{\prime }}^{-}\right\rangle =0$.
When a small randomly perturbation $H^{\prime }$\ is induced, the
eigenstates set becomes $\left\{ \left\vert \phi _{n}^{\sigma }\right\rangle
\right\} $\ with $\sigma =\pm $ and $n\in \lbrack 1,N/2]$.\ For sufficient
small perturbation, we always have\ $\langle \psi _{k}^{\sigma }\left\vert
\phi _{n}^{-\sigma }\right\rangle \approx 0,$which is simple but crucial to
this work. In a disordered system, $\left\{ \gamma _{ij}\right\} $\ becomes
a set of random numbers around the $\left\{ \gamma _{ij}^{0}\right\} $\ of $%
H_{0}$. In the large limit of disorder, the energy gap collapses and $%
\langle \psi _{k}^{\sigma }\left\vert \phi _{n}^{-\sigma }\right\rangle $\
cannot be neglected. Fig. \ref{fig1} schematically illustrates this point.

In the case that the perturbation does not collapse the original energy
band, state $\left\vert \phi _{n}^{\sigma }\right\rangle $ is only the
superposition of the sub-set $\left\{ \left\vert \psi _{k}^{\sigma
}\right\rangle \right\} $ approximately.\ In the following we focus on a
single band and neglect the band index $\sigma $. For an arbitrary operator $%
\mathcal{O}$, the matrix representation with $\left\{ \left\vert \psi
_{k}\right\rangle \right\} $\ is a $\left\langle \psi _{k}\right\vert
\mathcal{O}\left\vert \psi _{k^{\prime }}\right\rangle $. For a perturbed
system, the matrix representation of $\mathcal{O}$\ changes but leaves the
trace of the matrix unchanged approximately. Actually, the quasi-closed
mapping admits
\begin{equation}
\left\vert \phi _{n}\right\rangle =\sum_{k}c_{n}^{k}\left\vert \psi
_{k}\right\rangle ,
\end{equation}%
where $c_{n}^{k}=\langle \psi _{k}\left\vert \phi _{n}\right\rangle $,
satisfying $\sum_{k}c_{n^{\prime }}^{k}\left( c_{n}^{k}\right) ^{\ast
}=\delta _{nn^{\prime }}$ and $\sum_{n}c_{n}^{k}\left( c_{n}^{k^{\prime
}}\right) ^{\ast }=\delta _{kk^{\prime }}$. Then we have
\begin{equation}
\sum_{n}\left\langle \phi _{n}\right\vert \mathcal{O}\left\vert \phi
_{n}\right\rangle =\sum_{k}\left\langle \psi _{k}\right\vert \mathcal{O}%
\left\vert \psi _{k}\right\rangle .
\end{equation}%
In physics, we are interested in the quantity
\begin{equation}
C_{i,j}=\sum_{n}\left\langle \phi _{n}\right\vert c_{i}^{\dag
}c_{j}\left\vert \phi _{n}\right\rangle ,
\end{equation}%
which is termed as band correlation function (BCF) and has evident physical
significance. For example, $C_{j,j}$\ relates to particle density at $j$\
site. We know that $C_{j,j}$\ for $H_{0}$\ is identical when $j$ belongs to
a sublattice due to the translational symmetry, but it is a little
surprisingly that this feature still holds when a slight randomly
perturbation is added although the translational symmetry is broken. On the
other hand, a BCF characterizes the feature of the band, probably involving
the topology of the band.

\begin{figure*}[htbp]
\centering
\subfigure{
\begin{minipage}[b]{0.32\linewidth}
\includegraphics[width=1\linewidth]{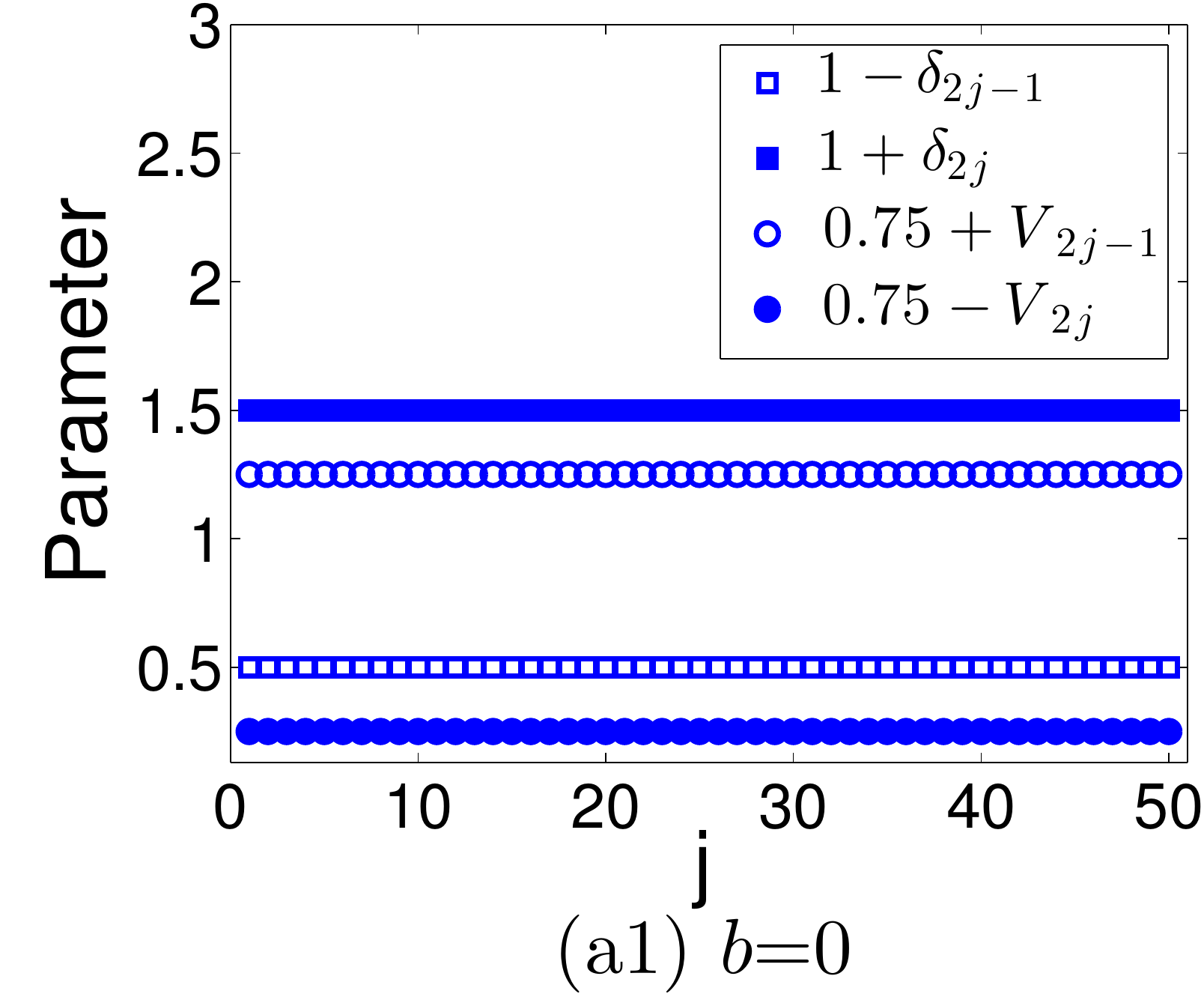}
\includegraphics[width=1\linewidth]{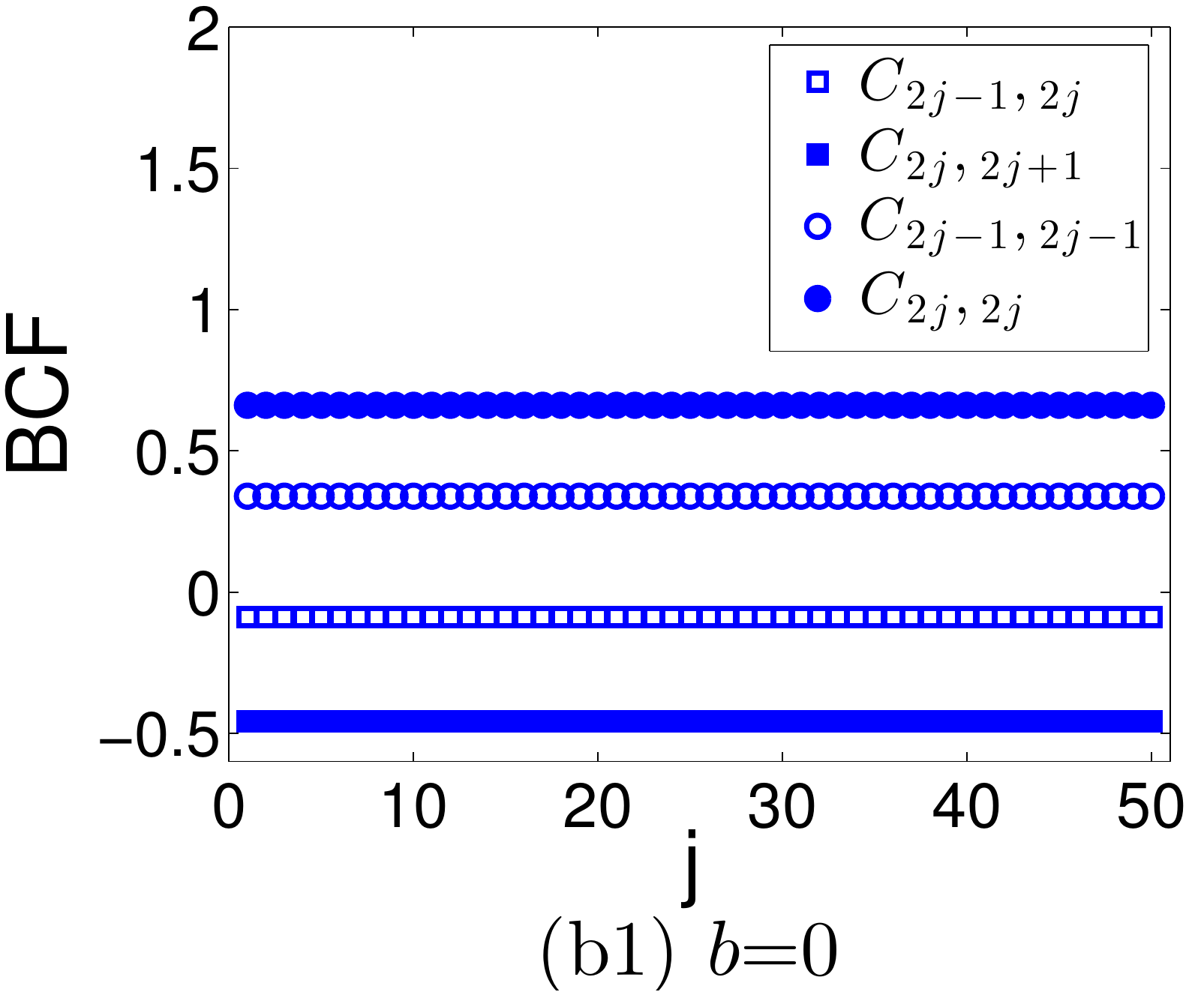}
\end{minipage}}
\subfigure{
\begin{minipage}[b]{0.32\linewidth}
\includegraphics[width=1\linewidth]{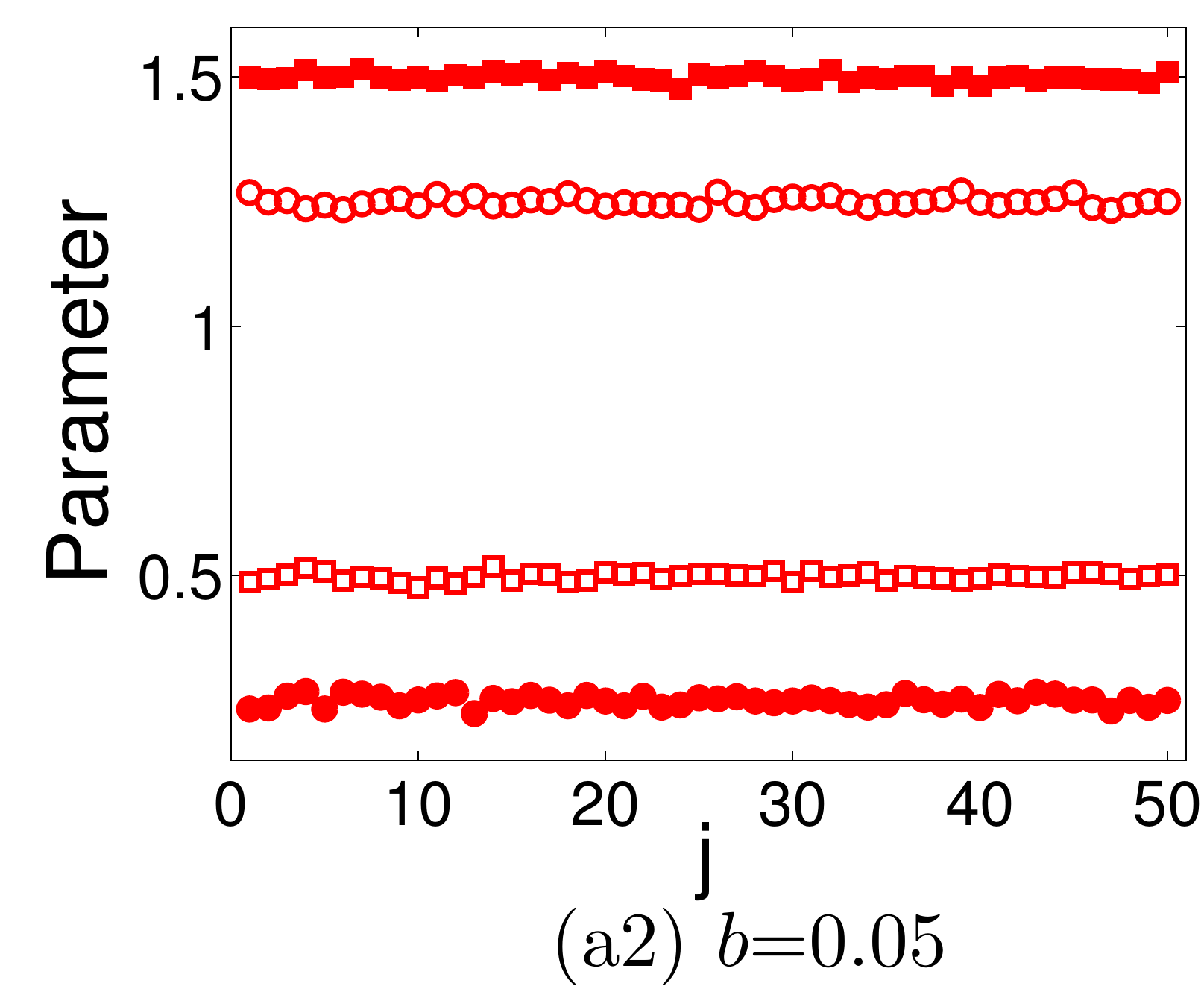}
\includegraphics[width=1\linewidth]{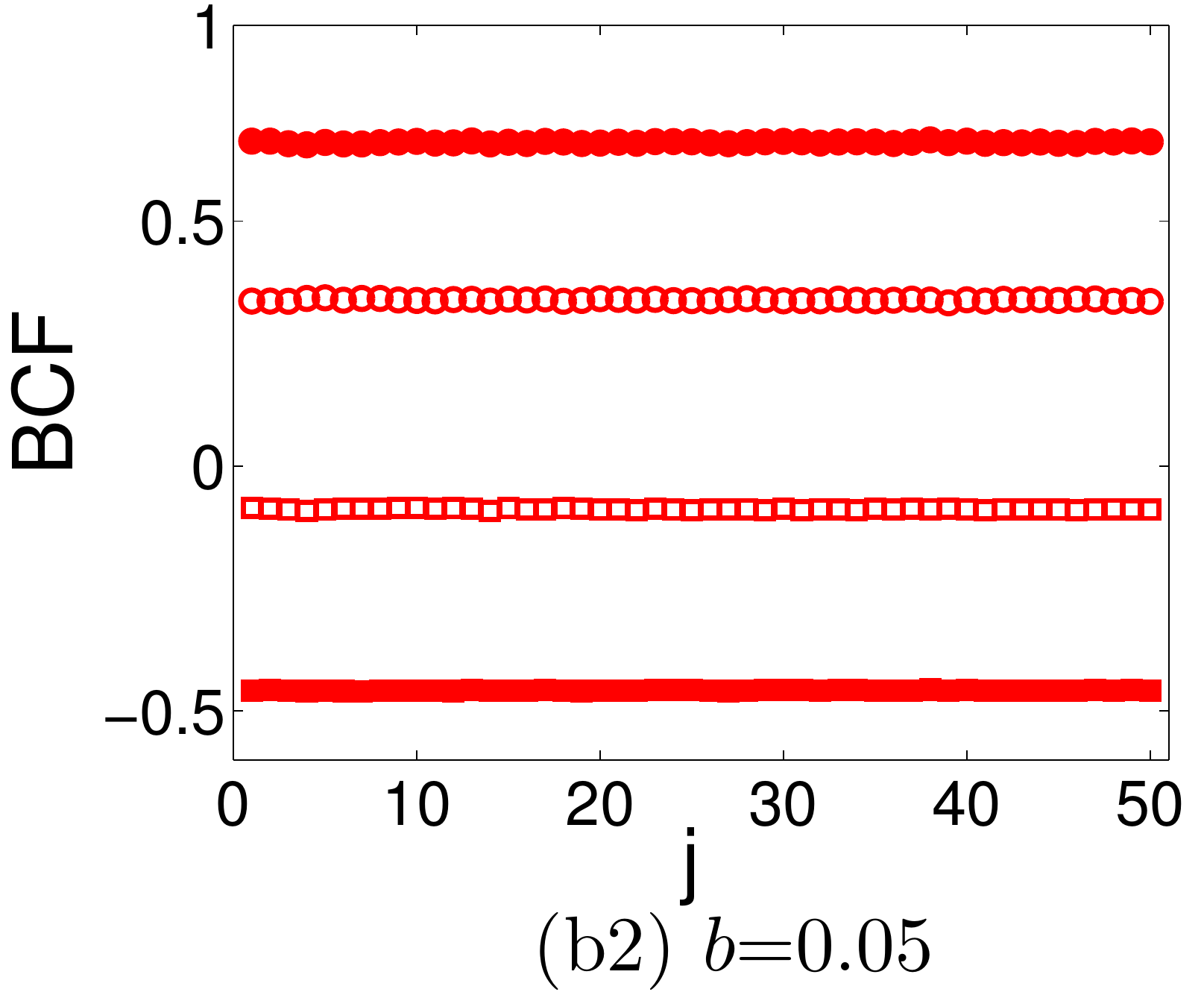}
\end{minipage}}
\subfigure{
\begin{minipage}[b]{0.32\linewidth}
\includegraphics[width=1\linewidth]{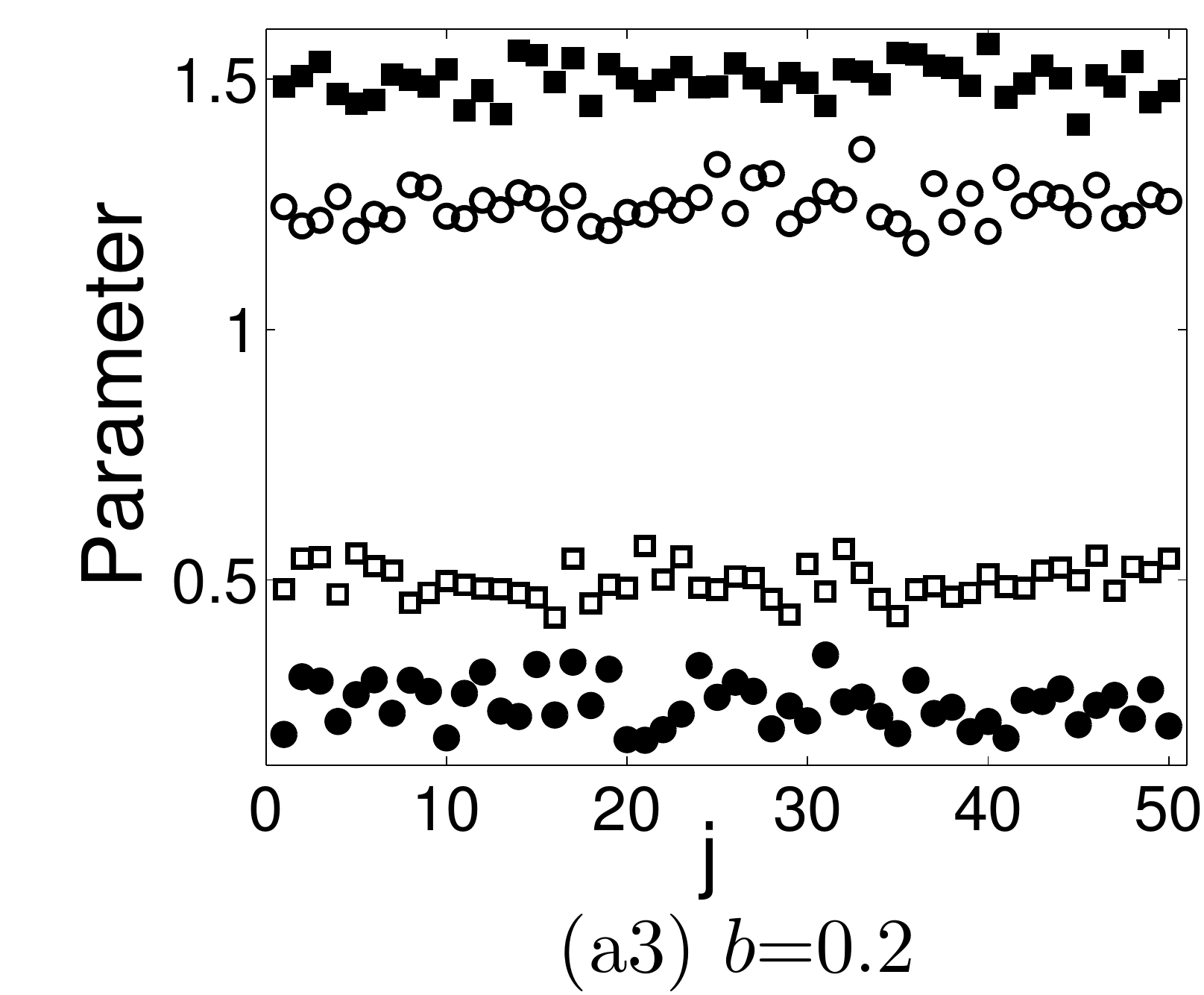}
\includegraphics[width=1\linewidth]{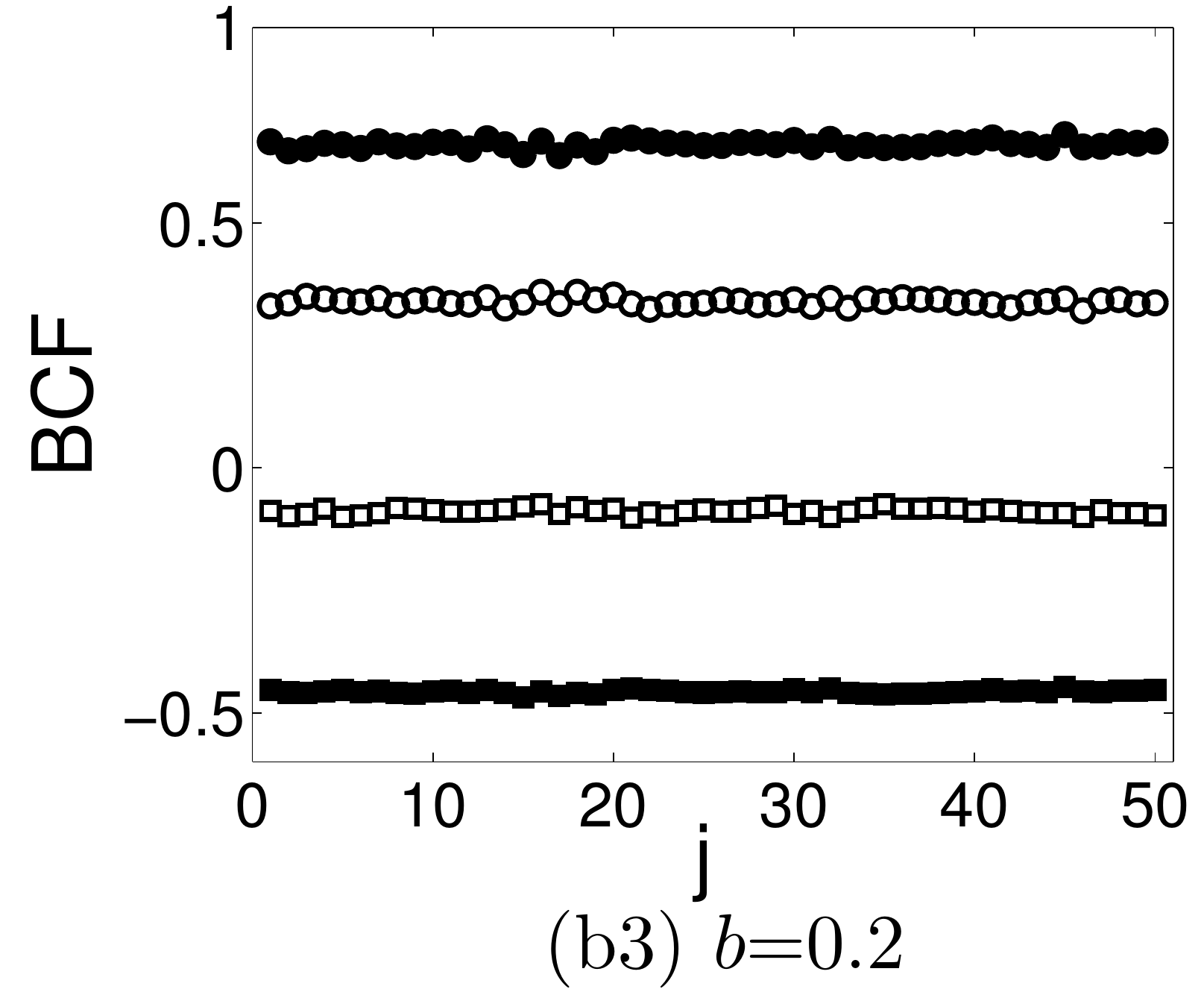}
\end{minipage}}
\caption{(Color online) Numerical simulations for RM models with different
strength of random parameters $(-b,b)$. (a1-a3) Plots of the hopping
constant and on-site potential distributions along the chain for three
typical random strengths. (b1-b3) Plots of the corresponding BCFs. The
system parameters are $N=50$, $V=0.5$, and $\protect\delta =0.5$. The small
random $b=0.05$\ almost does not affect the uniform distribution of four
types of BCFs. Surprisingly, as $b$ increases to $0.2$, the induced
fluctuations in (b3) are not so much in comparison with that in (a3).}
\label{fig2}
\end{figure*}

\begin{figure}[htbp]
\includegraphics[width=0.8\textwidth]{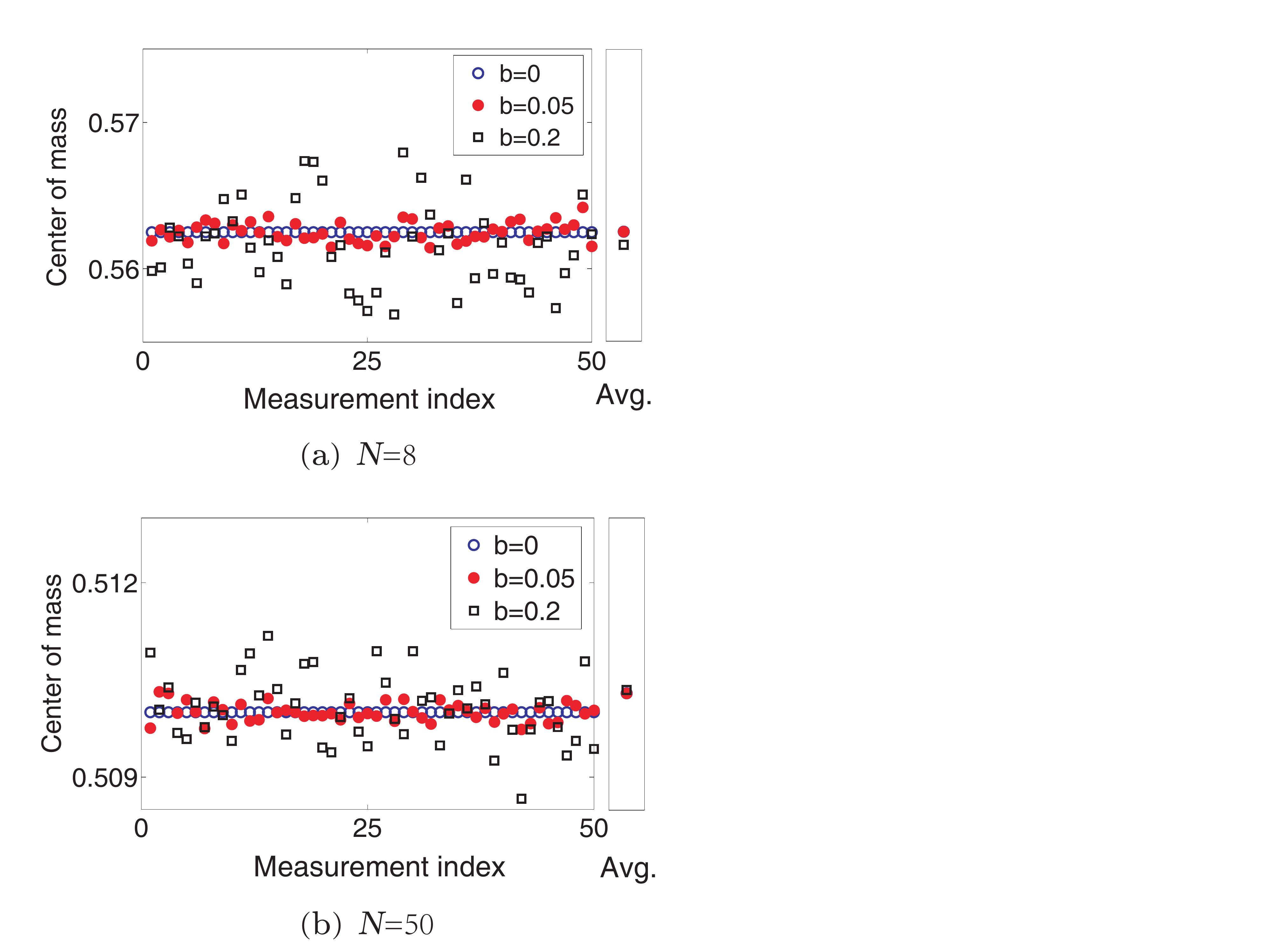}

\caption{(Color online) Plot of $\mathcal{M}_{c}$ (left panels) defined in
Eq. (\protect\ref{Mc2}) for disordered RM model with a sequence of random
number indicated by measurement index. The simulation is performed for
different random strengths with $b=0$ (empty blue circle), $b=0.05$ (solid
red circle) and $b=0.2$ (empty black square). The average of $\mathcal{M}%
_{c} $ (right panels) over $50$ times measurement shows that the robustness
of the center of mass for relatively large $b$ value. The length of the ring
(a) $N=8$, (b) $N=50$ and the parameters are $V=0.5$, $\protect\delta =0.5$.
Note that (a) and (b) are plotted in different scales, indicating that the
fluctuation of $\mathcal{M}_{c}$\ decreases as $N$ increases. In both cases,
the average of $\mathcal{M}_{c}$ are very close to that of zero $b$.}
\label{fig3}
\end{figure}

\begin{figure*}[htbp]
\includegraphics[width=1\textwidth]{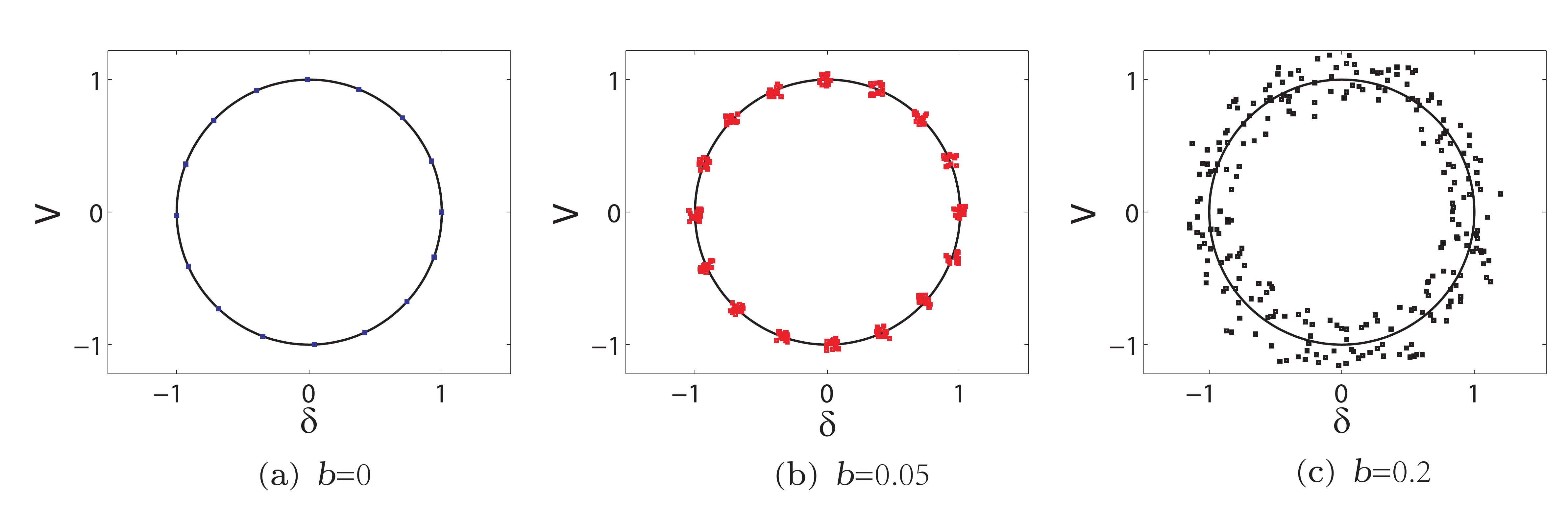}
\caption{(Color online) Plots of $\left\{ \protect\delta %
_{j}(t_{l}),V_{j}(t_{l})\right\} $\ in $V$-$\protect\delta $ plane from Eq. (%
\protect\ref{parameter}) with (a) $b=0$, (b) $b=0.05$, and (c) $b=0.2$. Here
we take $\protect\delta _{0}=V_{0}=0$, $R=1$, $M=16$ and $N=8$. The
corresponding discretized time is $t_{l}=2\protect\pi (l+1)/(16\protect%
\omega )$\ with \ $l=0,1,...,15$.}
\label{fig4}
\end{figure*}

\section{Disordered RM model}

\label{Disordered RM model}

As an example, we consider a RM model with a time-dependent random
perturbation on a $2N$-site lattice. It provides a natural platform for
studying topological invariant directly through dynamics both in theoretical
\cite{XiaoDi,WR1,WR2} and experimental \cite{Marcos} perspectives. The
Hamiltonian is
\begin{eqnarray}
H(t) &=&-\sum_{j=1}^{2N}[1+(-1)^{j}\delta _{j}(t)]c_{j}^{\dag }c_{j+1}+%
\mathrm{H.c.}  \notag \\
&&+\sum_{j=1}^{2N}(-1)^{j}V_{j}(t)c_{j}^{\dag }c_{j},  \label{H}
\end{eqnarray}%
where particle operator obeys periodic boundary condition $c_{2N+1}=c_{1}$.
Parameters $\left\{ \delta _{j}(t)\right\} $\ and $\left\{ V_{j}(t)\right\} $%
\ are two sets of time- and position-dependent numbers. In the following, we
first establish the relation between the topological quantities, Zak phase
and pumping charge, and the correlation functions. Subsequently, we
investigate the influence of disorder perturbation on the corresponding
correlation functions numerically.

\subsection{Regular RM model}

In the absence of the disorder, the Hamiltonian becomes\ $H_{0}(t)$ by
taking $\delta _{j}=\delta (t)$\ and $V_{j}=V(t)$. Before proceeding, we
briefly recall the solution of $H_{0}$, which reads
\begin{equation}
H_{0}=-\sum_{j=1}^{2N}[1+(-1)^{j}\delta ]c_{j}^{\dag }c_{j+1}+\mathrm{H.c.}%
+V\sum_{j=1}^{2N}(-1)^{j}c_{j}^{\dag }c_{j}.
\end{equation}%
It can be obtained by introducing linear transformation
\begin{equation}
\left\{
\begin{array}{c}
\alpha _{k}=N^{-1/2}\sum\limits_{j=1}^{N}e^{-ikj}(e^{i\phi }\sin \frac{%
\theta }{2}c_{2j-1}-\cos \frac{\theta }{2}c_{2j}) \\
\beta _{k}=N^{-1/2}\sum\limits_{j=1}^{N}e^{-ikj}(e^{i\phi }\cos \frac{\theta
}{2}c_{2j-1}+\sin \frac{\theta }{2}c_{2j})%
\end{array}%
\right. ,
\end{equation}%
where the $k$-dependent parameters are defined by
\begin{equation}
\tan \phi =\frac{(1+\delta )\sin k}{(1-\delta )+(1+\delta )\cos k},\tan
\theta =\frac{|\gamma _{k}|}{-V}.
\end{equation}%
Based on this, we have the diagonal form of the Hamiltonian
\begin{equation}
H_{0}=\sum_{k}\varepsilon _{k}(\alpha _{k}^{\dag }\alpha _{k}-\beta
_{k}^{\dag }\beta _{k}),
\end{equation}%
where the spectrum
\begin{equation}
\varepsilon _{k}=\sqrt{V^{2}+|\gamma _{k}|^{2}},
\end{equation}%
with $\gamma _{k}=(1-\delta )+(1+\delta )e^{ik}$\ and the momentum $k=2n\pi
/N,n=1,2,...,N$. The single-particle eigenvectors are $|\psi _{k}^{+}\rangle
=\alpha _{k}^{\dag }|0\rangle $ and $|\psi _{k}^{-}\rangle =\beta _{k}^{\dag
}|0\rangle $. In the following, we focus on the lower band by taking $|\psi
_{k}\rangle =|\psi _{k}^{-}\rangle $, the result for upper band can be
obtained directly.

The Zak phase as a related topological quantity can be expressed as
\begin{equation}
\mathcal{Z}=\frac{i}{2\pi }\int_{-\pi }^{\pi }\left\langle \psi
_{k}\right\vert \partial _{k}|\psi _{k}\rangle \mathrm{d}k=\mathcal{Z}_{%
\mathrm{ad}}+\mathcal{Z}_{\mathrm{com}},  \label{Z1}
\end{equation}%
where $\mathcal{Z}_{\mathrm{ad}}$\ attributes to adiabatic current and $%
\mathcal{Z}_{\mathrm{com}}$\ arises from the center of mass. Here we would
like to express them in term of the correlation functions%
\begin{eqnarray}
&&\mathcal{Z}_{\mathrm{ad}}=\frac{N}{2\pi }\int \left\langle \psi
_{k}\right\vert c_{2j-1}^{\dag }c_{2j-1}|\psi _{k}\rangle \mathrm{d}\phi -%
\frac{iN}{4\pi }\times  \notag \\
&&\int \left\langle \psi _{k}\right\vert (e^{-i\phi }c_{2j-1}^{\dag
}c_{2j}-e^{i(\phi -k)}c_{2j}^{\dag }c_{2j+1})|\psi _{k}\rangle \mathrm{d}%
\theta ,
\end{eqnarray}%
and%
\begin{equation}
\mathcal{Z}_{\mathrm{com}}=M_{c}=\sum_{j=1}^{N}jm_{j},  \label{Mc1}
\end{equation}%
where the weight at $j$th site%
\begin{equation}
m_{j}=\frac{1}{2N\pi }\int_{-\pi }^{\pi }\left\langle \psi _{k}\right\vert
(c_{2j-1}^{\dag }c_{2j-1}+c_{2j}^{\dag }c_{2j})|\psi _{k}\rangle \mathrm{d}k.
\end{equation}%
It indicates that the correlations $\left\langle \psi _{k}\right\vert
c_{j}^{\dag }c_{j}|\psi _{k}\rangle $\ and\ $\left\langle \psi
_{k}\right\vert c_{j}^{\dag }c_{j+1}|\psi _{k}\rangle $\ have intimate
connection with the Zak phase $\mathcal{Z}$.

Based on the the relations%
\begin{equation}
\left\{
\begin{array}{l}
\left\langle \psi _{k}\right\vert c_{2j-1}^{\dag }c_{2j-1}|\psi _{k}\rangle
=N^{-1}\cos ^{2}(\frac{\theta }{2}) \\
\left\langle \psi _{k}\right\vert c_{2j}^{\dag }c_{2j}|\psi _{k}\rangle
=N^{-1}\sin ^{2}(\frac{\theta }{2}) \\
\left\langle \psi _{k}\right\vert c_{2j-1}^{\dag }c_{2j}|\psi _{k}\rangle =%
\frac{1}{2}N^{-1}\sin \theta e^{i\phi } \\
\left\langle \psi _{k}\right\vert c_{2j}^{\dag }c_{2j+1}|\psi _{k}\rangle =%
\frac{1}{2}N^{-1}\sin \theta e^{i(k-\phi )}%
\end{array}%
\right. ,
\end{equation}%
we obtain the compact expressions
\begin{equation}
\mathcal{Z}_{\mathrm{ad}}=\frac{1}{2\pi }\int_{\phi }\cos ^{2}(\frac{\theta
}{2})d\phi
\end{equation}
and
\begin{equation}
\mathcal{Z}_{\mathrm{com}}=(N+1)/(2N).
\end{equation}%
The constancy of $\mathcal{Z}$\ allows
\begin{equation}
\mathrm{d}\mathcal{Z}=-\frac{1}{2\pi }\int_{\phi }\Omega _{\theta \phi }%
\mathrm{d}\phi \mathrm{d}\theta =J\mathrm{d}t,
\end{equation}%
where $\Omega _{\theta \phi }=\frac{1}{2}\sin \theta $\ is Berry curvature
and
\begin{equation}
J=\frac{i}{2\pi }\int_{-\pi }^{\pi }[(\partial _{t}\langle \psi
_{k}|)\partial _{k}|\psi _{k}\rangle -(\partial _{k}\langle \psi
_{k}|)\partial _{t}|\psi _{k}\rangle ]\mathrm{d}k
\end{equation}%
is current across two neighboring sites. For a loop with
\begin{equation}
\delta (t+T)=\delta (t),V(t+T)=V(t)
\end{equation}%
around the point $\delta _{c}=V_{c}=0$, the pumping charge $Q=\int_{0}^{T}J%
\mathrm{d}t=\pm 1$ as a demonstration of Chern number. It is presumably that
as a local quantity, $Q$\ should not change drastically in the presence of
perturbation and takes the value around the original integral. This may be
in accordance with the robustness of BCFs, which will be investigated
numerically in the following section.

\subsection{Disorder perturbation}

In the presence of disorder, the definition of Zak phase is undefined due to
the translational symmetry breaking. However, the concept of pumping charge
is independent of the translational symmetry. It turns out that one of two
ingredients of the pumping charge is the center of mass, which may not be a
constant in the presence of disorder perturbation. In this section, we
investigate the influence of disorder on the center of mass. In general, the
center of mass for a energy band is defined as
\begin{equation}
\mathcal{M}_{c}=N^{-2}\sum_{n,j=1}^{N}j\left\langle \phi _{n}\right\vert
(c_{2j-1}^{\dag }c_{2j-1}+c_{2j}^{\dag }c_{2j})|\phi _{n}\rangle ,
\label{Mc2}
\end{equation}%
where $\left\{ |\phi _{n}\rangle \right\} $ is the eigenstate of $H$ in Eq. (%
\ref{H}). When the perturbation switches on, i.e., $\delta _{j}\neq \delta $%
\ and $V_{j}\neq V$, the translational symmetry is broken. However,
according to our analysis in section \ref{Robustness of band correlation
function}, the BCFs $C_{i,j}=\sum_{n}\left\langle \phi _{n}\right\vert
c_{i}^{\dag }c_{j}\left\vert \phi _{n}\right\rangle $ are still robust if
the band gap does not collapse and we can have a conclusion with
\begin{eqnarray}
&&\sum_{n}^{N}\left\langle \phi _{n}\right\vert (c_{2j-1}^{\dag
}c_{2j-1}+c_{2j}^{\dag }c_{2j})|\phi _{n}\rangle  \notag \\
&\approx &\sum_{k}\left\langle \psi _{k}\right\vert (c_{2j-1}^{\dag
}c_{2j-1}+c_{2j}^{\dag }c_{2j})|\psi _{k}\rangle =1.
\end{eqnarray}%
Then $\mathcal{M}_{c}$\ is approximately a constant in the presence of
disorder. To demonstrate this point and find the extent to which the
deviation of $C_{i,j}$ is negligible, we numerically compute $C_{i,i}$ and $%
C_{i,i+1}$ for disorder RM model, in which $\left\{ \delta _{j}\right\} $\
and $\left\{ V_{j}\right\} $\ are two sets of random numbers around $\delta $%
\ and $V$, respectively. Actually, $C_{i,i}$\ and $C_{i,i+1}$\ are immune to
the random perturbation no matter $\left\{ \delta _{j}\right\} $\ and $%
\left\{ V_{j}\right\} $\ are real or imaginary.\ For clarity, we only
consider the real random numbers in later discussions.

Numerical simulation is performed by taking two sets of random numbers $%
\left\{ \delta _{j}\right\} $\ and $\left\{ V_{j}\right\} $ around $\delta $%
\ and $V$.\textbf{\ }The random number parameter can be taken as
\begin{equation}
\delta _{j}=\delta +\mathrm{ran}(-b,b),V_{j}=V+\mathrm{ran}(-b,b),
\end{equation}%
where $\mathrm{ran}(-b,b)$\ denotes a uniform random number within $(-b,b)$.
In Fig. \ref{fig2}, we plot the results with several typical values of $b$.
We find that $C_{i,i}$ and $C_{i,i+1}$ are still immune to the random
perturbation for relatively large $b$ value. This will result in the
robustness of $\mathcal{M}_{c}$\ against the disorder. To estimate the
effect of disorder, the fluctuation of $\mathcal{M}_{c}$, we plot the center
of mass with several typical values of $b$ in Fig. \ref{fig3}. It shows that
the fluctuation of $\mathcal{M}_{c}$\ is within $0.1\%$ for large size
system.\textbf{\ }The result indicates that although the translational
symmetry is broken, the center of mass is still time-independent
approximately, contributing much less to the current or pumping charge. It
is expected that the pumping charge can still take the role of topological
invariant.

\begin{figure*}[htbp]
\includegraphics[width=1\textwidth]{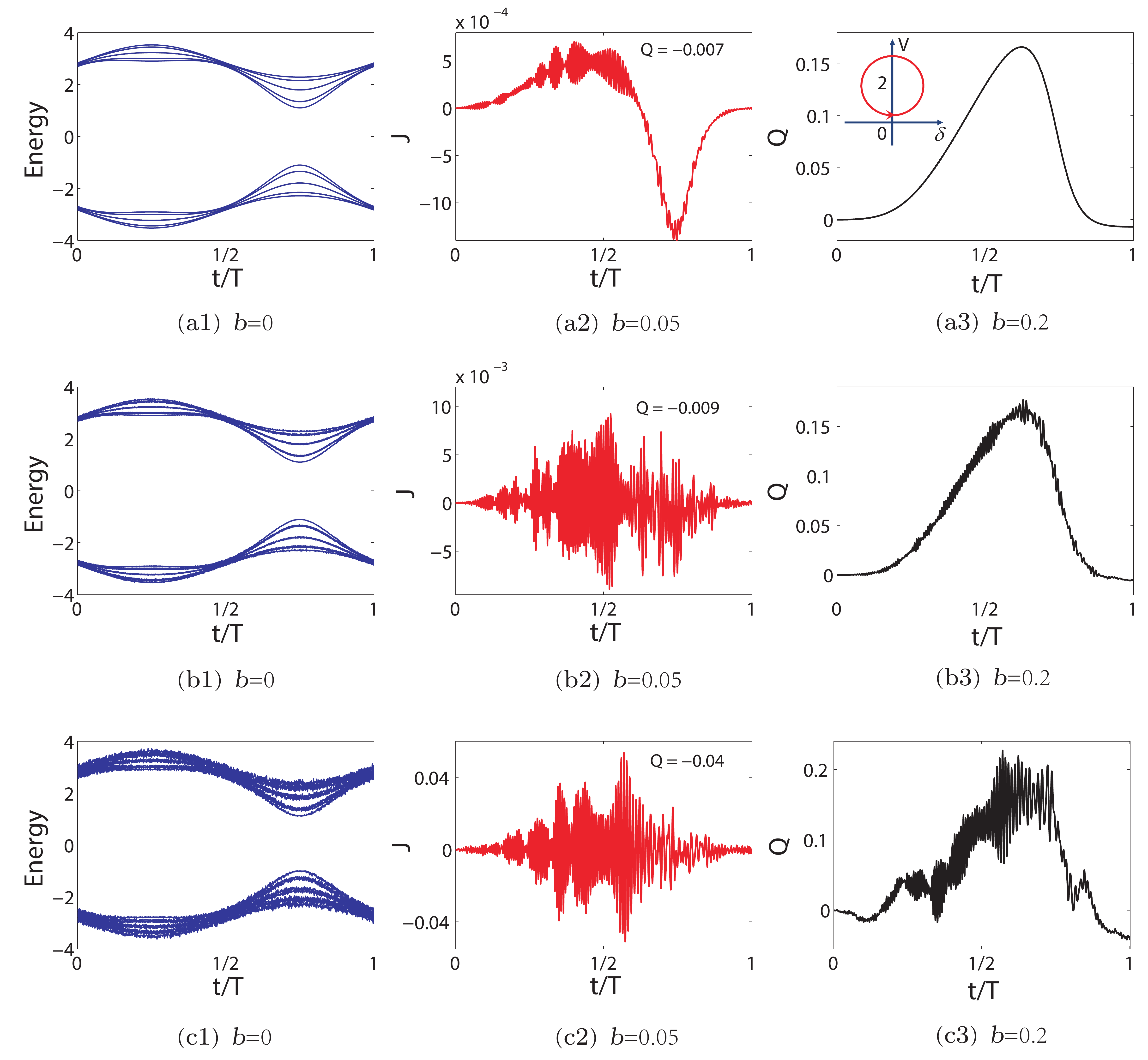}
\caption{(Color online) Numerical simulations of time evolution driven by
the time-dependent Hamiltonian with parameters in Eq. (\protect\ref%
{parameter}). Plots of the instantaneous spectra (a1-c1) local current
(a2-c2) and charge accumulation (a3-c3) as function of evolution time for
several typical random strengths $b$. The parameters are taken as $(\protect%
\delta _{0},V_{0})=(0,2)$, $R=1$, $N=8$, $\protect\omega =6\times 10^{-5}$,
and $M=5\times 10^{4}$. We note that the loop of time evolution does not
enclose the degeneracy point $(0,0)$. The obtained pumping charges are very
close zero, even for $b=0.2$, indicating the robustness of the pumping
charge.}
\label{fig5}
\end{figure*}

\begin{figure*}[htbp]
\includegraphics[width=1\textwidth]{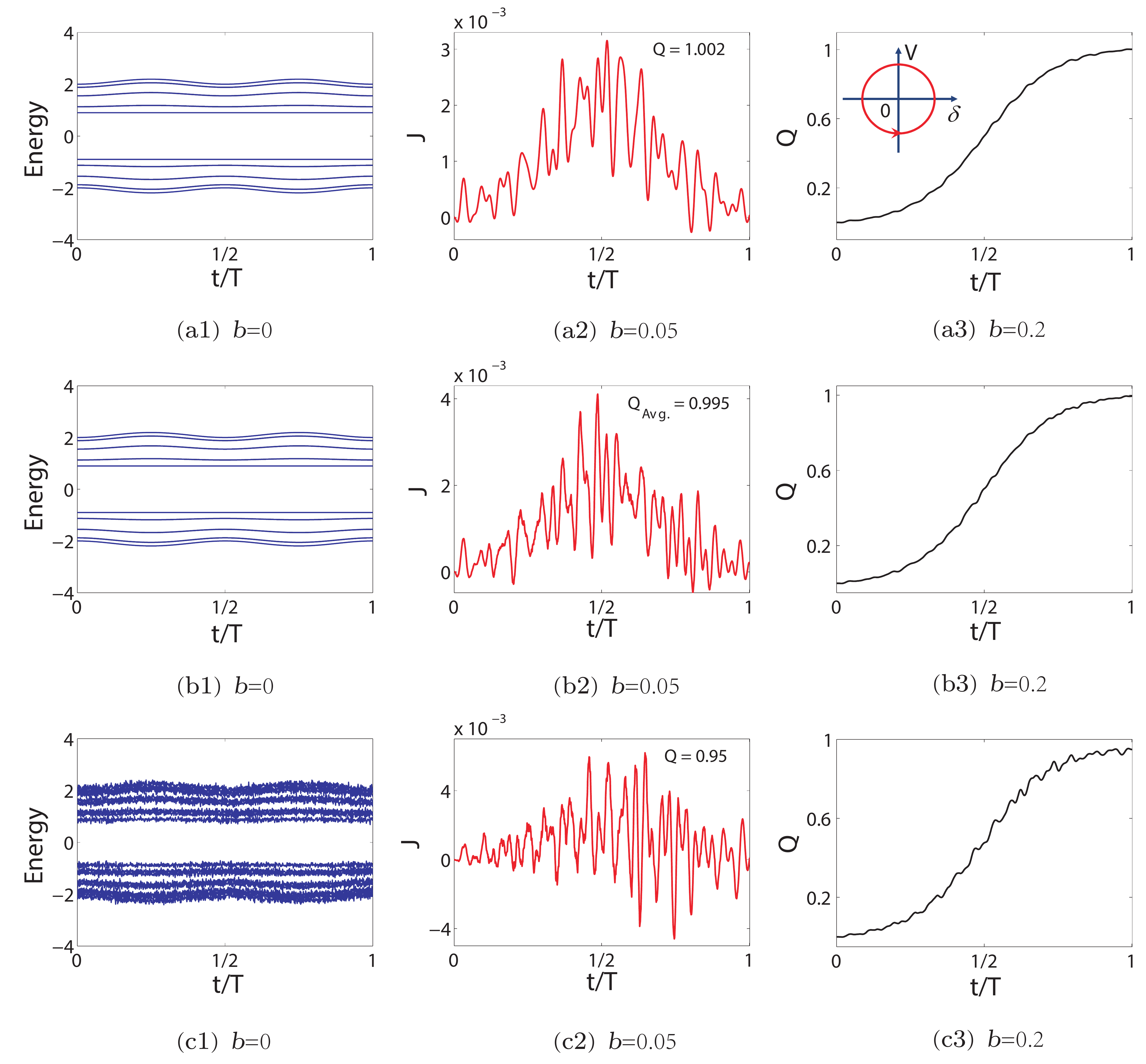}
\caption{(Color online) The same as Fig. \protect\ref{fig5} but for $(%
\protect\delta _{0},V_{0})=(0,0)$. The loop of time evolution encloses the
degeneracy point $(0,0)$. The obtained pumping charges is very close to $1$
for $b=0.05$, indicating the robustness of the pumping charge. In the case
with $b=0.2$, the pumping charge is close to $1$ with $5\%$ deviation.}
\label{fig6}
\end{figure*}

\begin{figure*}[tbph]
\includegraphics[width=1\textwidth]{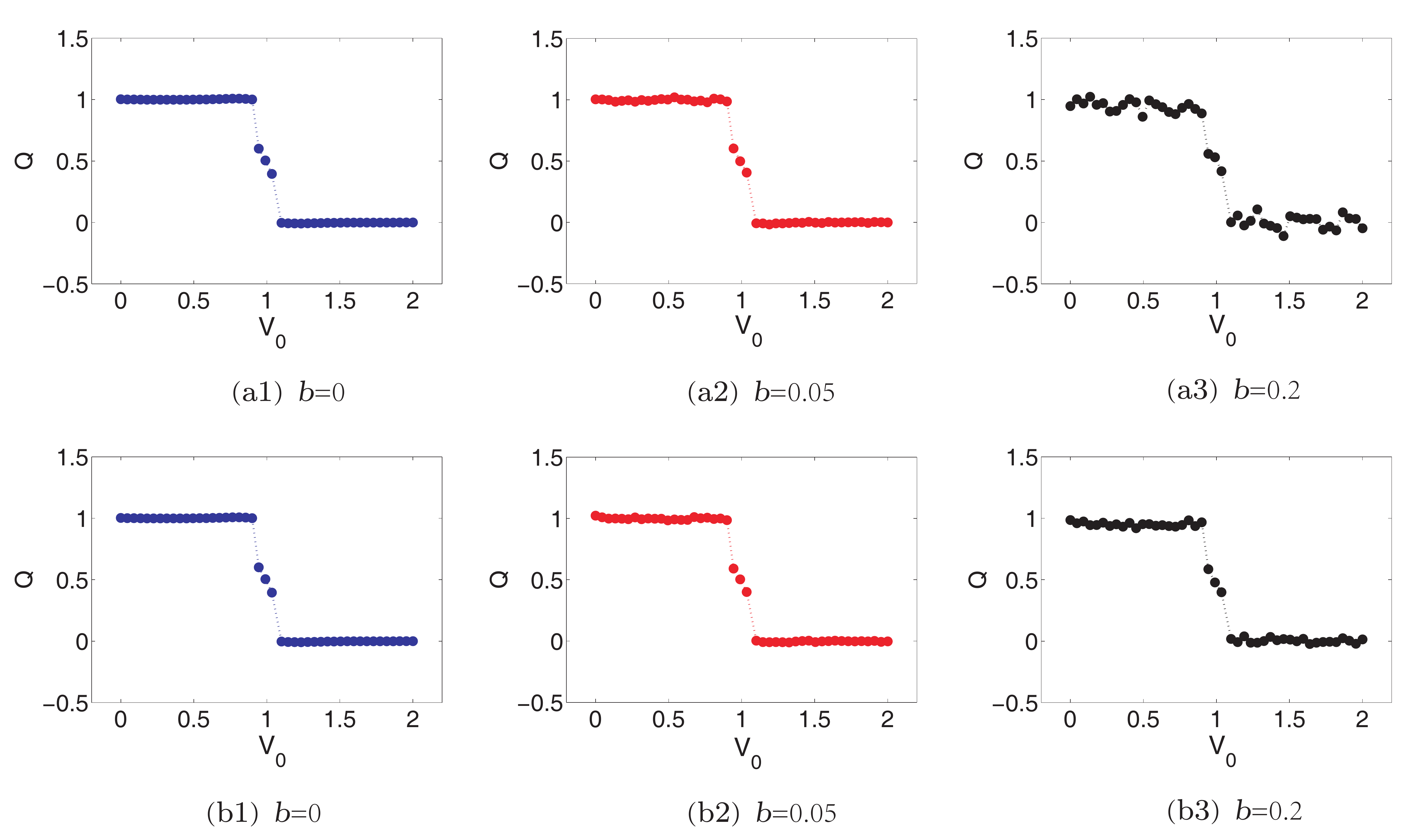}
\caption{(Color online) Plots of the pumping charge as function of the
position $(\protect\delta _{0},V_{0})=(0,V_{0})$, obtained by numerical
simulations as the same in Fig. \protect\ref{fig5} and Fig. \protect\ref%
{fig6} for systems with (a1-a3) $N=8$, (b1-b3) $N=50$. It indicates that $Q$%
\ experiences evident jump at the critical point, even for relatively large
random strength $b=0.2$. Large $N$ can increase the robustness of $Q$
against the disorder, supporting the concept of topological invariant in an
approximate manner.}
\label{fig7}
\end{figure*}

\section{Robustness of pumping charge}

\label{Robustness Of Pumping Charge}

In this section, we focus on the Thouless quantum pump in the disorder
system. Inspired by the above analysis, we expect that there exists
residual\ topological feature in the disordered RM model, which may be
unveiled by the pumping charge. Numerical simulation is performed by taking
\begin{eqnarray}
\delta _{j}(t) &=&\delta _{0}+R\cos (\omega t)+\mathrm{ran}(-b,b)_{t},
\notag \\
V_{j}(t) &=&V_{0}+R\sin (\omega t)+\mathrm{ran}(-b,b)_{t},
\end{eqnarray}%
where the subscript index $t$ indicates the random number is time dependent.
The computation is performed by using a uniform mesh in the time
discretization. The time evolution is a little different from the previous
works, where parameters vary smoothly in time. This is implemented by
discretizing the time $t$ into $t_{i}$, with $t_{0}=0$ and $t_{M}=T$. For a
given initial eigenstate $\left\vert \phi _{n}\left( 0\right) \right\rangle $%
, the time evolved state is computed by
\begin{equation}
\left\vert \phi _{n}\left( t_{l}\right) \right\rangle
=\prod\limits_{l=1}^{M}\exp [(-iH\left( t_{l-1}\right)
(t_{l}-t_{l-1})]\left\vert \phi _{n}(0)\right\rangle ,
\end{equation}%
where each $H\left( t_{l}\right) $ corresponds to a set of random number $%
\mathrm{ran}(-b,b)$. In this work, the parameters for $H\left( t_{l}\right) $%
\ is taken in the form
\begin{eqnarray}
\delta _{j}(t_{l}) &=&\delta _{0}+R\cos (\omega t_{l})+\mathrm{ran}%
(-b,b)_{l},  \notag \\
V_{j}(t_{l}) &=&V_{0}+R\sin (\omega t_{l})+\mathrm{ran}(-b,b)_{l},
\label{parameter}
\end{eqnarray}%
with $l\in \left[ 0,M-1\right] $. In Fig. \ref{fig4}, we illustrate this
scheme by plotting $\left\{ \delta _{j}(t_{l}),V_{j}(t_{l})\right\} $\ with
several typical $b$ but a small $M$ and $N$. In the simulation, the number
of $M$\ is taken to be sufficient large in order to get convergent result.
The total pumping charge passing the site $j$ attributed from all channels
of lower energy band during the time evolution period $T$, can be expressed
as
\begin{eqnarray}
Q &=&\sum_{n=1}^{N}\int_{0}^{T}\langle \phi _{n}(t)|\mathcal{J}%
_{j}\left\vert \phi _{n}(t)\right\rangle \mathrm{d}t  \notag \\
&\approx &\sum_{n=1}^{N}\sum_{l=1}^{M}\langle \phi _{n}\left( t_{l}\right) |%
\mathcal{J}_{j}\left\vert \phi _{n}\left( t_{l}\right) \right\rangle
(t_{l}-t_{l-1}),
\end{eqnarray}%
which is independent of $j$\ for an adiabatic circle. Here the current
operator is
\begin{equation}
\mathcal{J}_{j}=\frac{1}{i}\left\{
\begin{array}{c}
(1-\delta _{j})|j\rangle \langle j+1|-\mathrm{H.c.}\mathbf{,}(j=2m-1) \\
(1+\delta _{j})|j\rangle \langle j+1|-\mathrm{H.c.}\mathbf{,}(j=2m)%
\end{array}%
\right. ,
\end{equation}%
where $m=1,2,...,N$. Fig. \ref{fig5} and Fig. \ref{fig6} plot the
simulations of particle current and the corresponding total probability for
several typical cases, in order to see to what extent the pumping charge can
be regarded as an quasi topological invariant. Fig. \ref{fig7} plots the
pumping charges for different loops and random strengths. It indicates that
weak disorder cannot affect the quantization of the pumping charge, and in
the case with strong disorder, pumping charge can still identify the
transition point approximately. As the size of the system increases, the
influence of disorder becomes weak. It shows that the pumping charge is a
reliable quantity to measure the topological invariant in experiment. Our
finding indicates that the topological feature may also emerge in a
noncrystalline system in an approximate manner.

\section{Summary}

\label{Summary}

In summary, we have analyzed the insensitivity of a topological invariant to
a disorder perturbation. In general, we find that there are a kinds of
quantities, such as BCFs, which are robust against the disorder. The
underlying mechanism is that the weak disorder cannot destroy the energy
band structure although it breaks the translational symmetry. This manifests
that a noncrystalline system may be topologically nontrivial in an
approximate manner, since an approximate topological invariant can be
retrieved. We have demonstrated this point by a time-dependent disordered RM
model. We computed the dynamical topological invariant by accumulating the
measurable local current without the requirement of translational symmetry.
The results indicate that (i) weak disorder almost has no effect on the
integer of Chern number; (ii) as disorder increases, the quasi Chern number
deviates from integer, but is still evidently to characterize the phase
transition. Our finding provides a way of extracting topological invariant
from imperfect system and predicts the existence of quasi topological phase.

\acknowledgments We acknowledge the support of National Natural Science
Foundation of China (Grant No. 11874225).

\end{document}